\documentclass[prb,aps,twocolumn,footinbib,floatfix,10pt,longbibliography,superscriptaddress]{revtex4-1}

\usepackage{chemformula} 
\usepackage[T1]{fontenc} 
\usepackage{graphicx}
\usepackage{dcolumn}
\usepackage{lipsum}
\usepackage{tabu} 
\usepackage{mathtools}

\usepackage{braket}
\usepackage{color}
\usepackage[11pt]{moresize}
\usepackage{anyfontsize}
\usepackage{bbding}
\usepackage{amsmath}
\usepackage{xcolor}
\usepackage{braket}
\usepackage[normalem]{ulem}

\begin{document}


\author{Patrick Laferri{\`e}re}
\email{plaferr3@uottawa.ca}
\affiliation{National Research Council Canada, Ottawa, Ontario, Canada, K1A 0R6.}
\affiliation{University of Ottawa, Ottawa, Ontario, Canada, K1N 6N5.}
\author{Aria Yin}
\affiliation{National Research Council Canada, Ottawa, Ontario, Canada, K1A 0R6.}
\affiliation{University of Ottawa, Ottawa, Ontario, Canada, K1N 6N5.}
\author{Edith Yeung}
\affiliation{National Research Council Canada, Ottawa, Ontario, Canada, K1A 0R6.}
\affiliation{University of Ottawa, Ottawa, Ontario, Canada, K1N 6N5.}
\author{Leila Kusmic}
\affiliation{National Research Council Canada, Ottawa, Ontario, Canada, K1A 0R6.}
\affiliation{University of Ottawa, Ottawa, Ontario, Canada, K1N 6N5.}
\author{Marek Korkusinski}
\affiliation{National Research Council Canada, Ottawa, Ontario, Canada, K1A 0R6.}
\author{Payman Rasekh}
\affiliation{National Research Council Canada, Ottawa, Ontario, Canada, K1A 0R6.}
\author{David B. Northeast}
\affiliation{National Research Council Canada, Ottawa, Ontario, Canada, K1A 0R6.}
\author{Sofiane Haffouz}
\affiliation{National Research Council Canada, Ottawa, Ontario, Canada, K1A 0R6.}
\author{Jean Lapointe}
\affiliation{National Research Council Canada, Ottawa, Ontario, Canada, K1A 0R6.}
\author{Philip J. Poole}
\affiliation{National Research Council Canada, Ottawa, Ontario, Canada, K1A 0R6.}
\author{Robin L. Williams}
\affiliation{National Research Council Canada, Ottawa, Ontario, Canada, K1A 0R6.}
\author{Dan Dalacu}
\affiliation{National Research Council Canada, Ottawa, Ontario, Canada, K1A 0R6.}
\affiliation{University of Ottawa, Ottawa, Ontario, Canada, K1N 6N5.}

\title{Approaching transform-limited photons from nanowire quantum dots excited above-band}


\begin{abstract}

We demonstrate that, even when employing above-band excitation, photons emitted from semiconductor quantum dots can have linewidths that approach their transform-limited values. This is accomplished by using quantum dots embedded in bottom-up photonic nanowires, an approach which mitigates several potential mechanisms that can result in linewidth broadening: (i) only a single quantum dot is present in each device, (ii) dot nucleation proceeds without the formation of a wetting layer, and (iii) the sidewalls of the photonic nanowire are comprised not of etched facets, but of epitaxially grown crystal planes. Using these structures we achieve linewidths of 2x the transform limit, unprecedented for above-band excitation. We also demonstrate a highly nonlinear dependence of the linewidth on both excitation power and temperature which can be described by an independent Boson model that considers both deformation and piezoelectric exciton-phonon coupling. We find that for sufficiently low excitation powers and temperatures, the observed excess broadening is not dominated by phonon dephasing, a surprising result considering the high phonon occupation that occurs with above-band excitation.

\end{abstract}

\maketitle 

Quantum interference between two or more indistinguishable photons lies at the heart of most photonic quantum technologies \cite{OBrien_NP2009} and high visibilities require highly coherent photons. Solid-state two-level emitters \cite{Aharonovich_NP2016}, e.g. epitaxial semiconductor quantum dots \cite{Senellart_NN2017}, can provide single photons with high efficiency provided they are incorporated within appropriate photonic structures supporting a single optical mode into which all photons are emitted. Within this solid-state environment, however, fluctuations in charge and spin or interactions with phonons can lead to broadening of the emitted photon linewidth compared to the Fourier transform limit and, consequently, a reduction of the two-photon interference visibility, $\nu_{\rm{TPI}}$ \cite{Vural_APL2020}. 

Excess broadening can be limited through the use of optical cavities \cite{Gerard_LT1999,Ding_PRL2016} and resonant excitation\cite{Muller_PRL2007}, resulting in the observation of $\nu_{\rm{TPI}}>90\%$ between sequential photons emitted from the same source over times scales $>1\mu \rm{sec}$ \cite{Wang_PRL2016,Tomm_NN2021}. Linewidths, typically measured over longer time scales, remain, however, broadened compared to the transform limit \cite{Wang_PRL2016,pulsed_vs_cw}, a broadening typically attributed to a slowly varying charge environment. One can therefore expect a reduction of $\nu_{\rm{TPI}}$ for times scales longer than a few microseconds and, in the case of remote emitters \cite{Patel_NP2010,Flagg_PRL2010,Gao_NC2013,Gold_PRB2014,Giesz_PRB2015,Reindl_NL2017,Weber_PRB2018} for which noise correlations are absent, the reduction may be particularly severe.

It is therefore essential to address this persistent broadening which arises from a time-dependent occupation of traps close to the dot, producing a wandering of the exciton energy via the Stark effect\cite{Robinson_PRB2000}. These traps can be defects or impurities located in either the bulk semiconductor material \cite{Gerardot_APL2011}, at surfaces (e.g. the growth surface \cite{Wang_APL2004} or the etched sidewall surface \cite{Liu_PRA2018}) or epitaxial interfaces \cite{Houel_PRL2012}. Additionally, quantum dots themselves are, by design, carrier traps and the impact on linewidth, at least of crystal phase dots \cite{Bavinck_NL2016} (i.e. stacking faults), is well established \cite{Dalacu_NT2019}. 

In this work we investigate a quantum dot system \cite{Laferriere_SR2022} distinct from the more conventional GaAs-based self-assembled dots \cite{Dietrich_LPR2016} in several aspects that impact excess broadening. The dots are segments of InAsP within [0001] wurtzite-phase InP nanowires and are fabricated using a bottom-up technique. Here the only surfaces in close proximity to the dot are the crystal lattice growth facets that make up the sidewalls of the nanowire, surfaces which have a lower defect density compared to the dry-etched sidewalls present in top-down structures \cite{Claudon_NP2010}. Using the InP material system, we can also expect an order of magnitude reduction in the surface recombination velocity compared to GaAs \cite{Hoffman_JAP1980}. The quantum dots are nucleated via a vapour-liquid-solid (VLS) growth process \cite{Borgstrom_NL2005} in which no 2D wetting layer forms, a layer associated with several broadening mechanisms \cite{Lobl_CP2018}. The VLS growth mode is also unique in that the number of dots in each device is controlled \cite{Laferriere_NL2020} and by using devices that contain only a single emitter, potential charge fluctuations associated with carriers trapped in nearby quantum dots are eliminated. 

We observe a significant reduction of excess broadening compared to conventional quantum dot emitters operated under the same conditions, with measured linewidths as low as 2x the transform limit. Remarkably, these narrow lines are observed using above-band excitation i.e. conditions expected to promote fluctuations in the charge environment. We also find that, at sufficiently low temperatures, broadening due to phonon dephasing is insignificant with an onset for strong broadening observed for $T>8$\,K. This is similar to that observed using resonant excitation\cite{Lobl_CP2018} and four-wave mixing techniques \cite{Borri_PRB2005} and consistent with a significant reduction in $\nu_{\rm{TPI}}$ observed in Ref.~\citenum{Reigue_PRL2017}. This behaviour can be fully described by an independent boson model \cite{Takagahara_PRB1999} that considers phonons propagating along the nanowire growth direction and which takes into account both deformation and piezoelectric exciton-phonon coupling. Importantly, we do not need to invoke virtual transitions between ground and excited states \cite{Muljarov_PRL2004,Reigue_PRL2017}, for which the energy scales are inconsistent with our dot level structure. Instead, the nonlinearity is traced to the interplay of deformation and piezoelectric coupling mechanisms and their distinct dependence on phonon energy.

\begin{figure}[h]
\includegraphics*[width=7.5cm,clip=true]{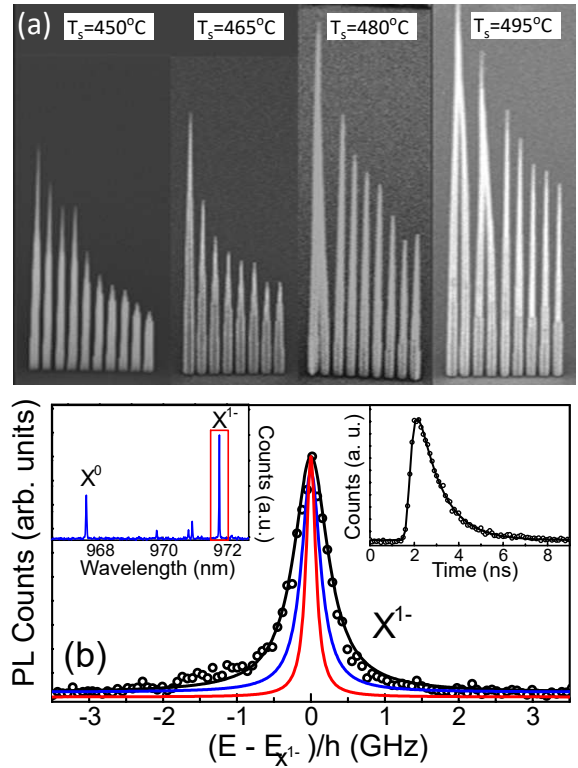}
\caption{
(a) Scanning electron microscope images of linear arrays of nanowires pitched at 400\,nm, each nanowire in an array having a different core diameter and each array grown at a different $T_s$, see Ref.~\onlinecite{Dalacu_APL2011} for details. (b) High resolution PL spectrum measured at 4\,K and $P=0.1P_{\mathrm{sat}}$ (black open circles) of the $X^{1-}$ line emission highlighted in the left inset. Also plotted are a Lorentzian fit to the data (black curve), the spectrum after deconvolution with the etalon response (blue curve) and the predicted transform-limited spectrum (solid red curve) calculated using the transition lifetime $T_1$ extracted from the PL decay curve shown in the right inset.}\label{fig_1}
\end{figure}

The nanowire-based devices were grown using a two-step growth process\cite{Dalacu_APL2011}. First, a 20\,nm diameter InP nanowire core is grown using growth conditions that promote axial growth (e.g. low Group V flux) and in this core we incorporate a $\sim 5$\,nm thick InAs$_{0.25}$P$_{0.75}$ quantum dot. Second, the core is clad with a shell using growth conditions that promote radial growth (high Group V flux). The resulting photonic nanowire has a base diameter of 250\,nm that tapers to $\sim 100$\,nm over a length of $\sim 15\,\mu$m, Figure~\ref{fig_1}(a). Previously, we have used growth temperatures of $420^\circ$C for both the core and shell growths which produced quantum dots emitting exciton photons with linewidths of 880\,MHz of which the inhomogeneous contribution was 730\,MHz\cite{Reimer_PRB2016} and was associated with a higher defect density in the InP causing charge fluctuations discussed above. In order to improve the material quality, in this study we increased the nanowire core growth temperature to $435^\circ$C, thought to be an upper limit above which reasonable axial growth rates could not be maintained \cite{Dalacu_NT2009}. For the shell growth we can access higher temperatures and here we study samples where the shell is grown at temperatures up to $T_\mathrm{s}= 495^\circ$C.

Photoluminescence (PL) measurements were performed in a closed-cycle helium cryostat. The nanowire quantum dots were excited through a 100x objective (numerical aperture of 0.81) and the emission was collected through the same objective and directed to a grating spectrometer equipped with a liquid nitrogen cooled charged-coupled device (CCD). For time-integrated PL measurements, the excitation was a continuous wave laser operating at 780\,nm whilst for time-resolved measurements, a pulsed laser operating at 670\,nm was used. To measure linewidths, the dot emission was coupled to a single mode fiber and a single peak was selected using a fiber-based tunable filter with a 0.1\,nm bandwidth. The spectral width of the filtered peak was resolved using a fiber-based piezo-driven scanning Fabry-P{\'e}rot etalon (bandwidth of 250\,MHz, free-spectral range of 40\,GHz) and detected with an avalanche photodiode. 

\begin{figure}[h]
\includegraphics*[width=8.cm,clip=true]{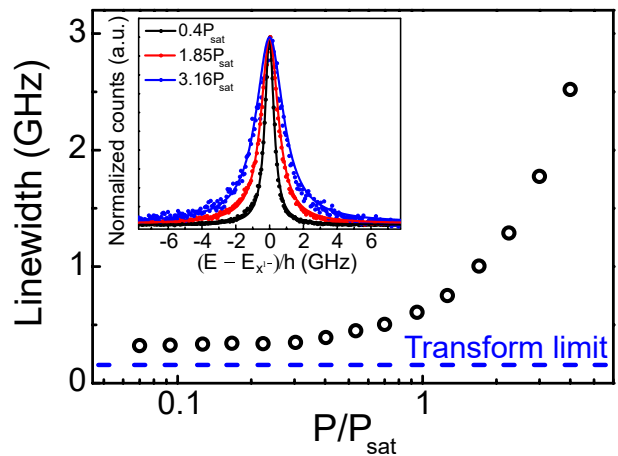}
\caption{Excitation power-dependence of the homogeneous linewidth extracted from the PL spectra after deconvolution with the etalon response function. Inset shows the high resolution PL spectra of the $X^{1-}$ emission measured at $T=4$\,K for selected excitation powers. Solid lines correspond to Lorentzian fits to the data.}\label{fig_2}
\end{figure}

We first look at the linewidth of an excitonic photon at low temperature (4\,K) and low excitation power ($P=0.1P_{\mathrm{sat}}$) in a device where the shell is deposited at $T_\mathrm{s}=450^\circ$C. Figure~\ref{fig_1} shows the high-resolution PL spectrum of an $X^{1-}$ photon emitting at $E_{X^{1-}}=1.276$\,eV. The emission peak was fit with a Lorentzian function to extract the homogeneous linewidth which, after deconvolution with the etalon response, is $\delta_\omega=322\,$MHz. We note that fitting with a Voigt profile produced a negligible Gaussian contribution (i.e. we did not observe any significant inhomogeneous broadening). Using the radiative recombination lifetime of $T_1=1.02 $\,ns from the PL decay curve (see inset in Figure~\ref{fig_1}), we calculate the width of the transform-limited spectrum from $\delta_\omega=1/2\pi T_1$. From these two widths we obtain the ratio $2T_1/T_2=2$, i.e. 2x the transform limit. This value is significantly improved compared to previous measurements of quantum dots using above-band excitation\cite{Birkedal_PRL2001,Bayer_PRB2002_1,Urbaszek_PRB2004,Berthelot_NP2006,Kuhlmann_NP2013,He_NN2013}, including our own nanowire devices grown at lower temperature \cite{Reimer_PRB2016}. It is also an improvement compared to measurements made using p-shell excitation \cite{Bayer_PRB2002_1,He_NN2013}. Instead, the values here are typically observed only when the emitter is excited resonantly\cite{Hogele_PRL2004,Atature_Sci2006,Houel_PRL2012,Kuhlmann_NP2013} and are not significantly greater than those observed in state of the art devices \cite{He_NN2013,Kuhlmann_NC2015,Wang_PRL2016}.

We note that further increases in $T_s$ did not result in any additional reduction of the linewidths suggesting that the excess broadening is not related to defects in the shell material \cite{comment_Tc}. To rule out broadening associated with phonon dephasing, we performed both power- and temperature-dependent measurements.  In Figure ~\ref{fig_2} we show the linewidth dependence on excitation power for powers up to $P=4P_{\mathrm{sat}}$. All power-dependent spectra show Lorentzian lineshapes (see inset) from which the homogeneous linewidths plotted in the figure are extracted. The dependence is highly nonlinear with the linewidth independent of excitation power for $P<0.4P_{\mathrm{sat}}$ above which strong broadening is observed. 
 
\begin{figure}[h]
\includegraphics*[width=8.cm,clip=true]{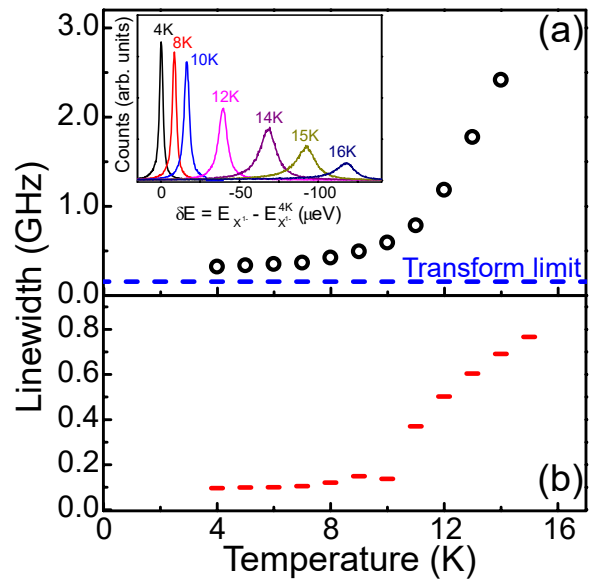}
\caption{(a) Deconvolved linewidth as a function of temperature in the low excitation power regime. Inset shows PL spectra, normalized to the integrated intensity, as a function of relative emission energy for selected temperatures. (b) Predicted broadening based on model described in text.}\label{fig_3}
\end{figure}

In Figure~\ref{fig_3} we show temperature-dependent measurements under weak pumping conditions ($P<0.2P_{\mathrm{sat}}$). With increasing temperature the emission peak red-shifts due to lattice dilation\cite{Varshni} and broadens. For temperatures up to $T=14\,K$ the lineshape remains Lorentzian whilst for $T>14\,K$ a slight asymmetry develops. We focus on the spectra measured for $T \leq 14\,K$ from which we can extract the homogeneous linewidths plotted in the figure. Significant broadening is observed for $T>8\,K$ whilst for $T<8\,K$, linewidths show only a small temperature dependence. We conclude that for sufficiently low temperatures ($T<8\,K$) and excitation powers ($P<0.4P_{\mathrm{sat}}$) phonon dephasing, notwithstanding above-band excitation, is not the source of the remaining excess broadening. Instead, it likely arises from some residual inhomogeneous broadening, that either does not sufficiently alter Lorentzian lineshapes such that it can be revealed by a simple lineshape fitting procedure or that manifests as a homogeneous broadening \cite{Berthelot_NP2006}. 

We next consider the nonlinear behaviour observed with temperature which is in stark contrast to the linear dependence seen in earlier studies\cite{comment_fat} that also employed non-resonant excitation \cite{Besombes_PRB2001,Birkedal_PRL2001,Bayer_PRB2002_1,Ortner_PRB2004,Urbaszek_PRB2004,Favero_PRB2007}. To date, the existence of the linear broadening is accounted for in InAs/GaAs zincblende dots by considering the independent boson model~\cite{Takagahara_PRB1999}, in which the exciton causes a renormalization of phonon modes without itself being excited by the lattice vibrations. The origin of the quadratic term, on the other hand, is traced to the virtual transitions between the ground and excited exciton states caused by the absorption and re-emission of the phonons\cite{Muljarov_PRL2004,Reigue_PRL2017}. In each of these approximations, only longitudinal phonon modes are accounted for, and the main exciton-phonon coupling mechanism is taken to be the deformation potential. In addition, the virtual transitions appear to be thermally activated only above a threshold temperature (of order of $10$\,K), below which only the linear broadening is expected\cite{Muljarov_PRL2004}. Within this general approach, it is straightforward to understand the similarity between the linewidth broadening as a function of excitation power (Figure~\ref{fig_2}) and as a function of temperature (Figure~\ref{fig_3}). Indeed, above-bandgap excitation creates electron-hole pairs in the InP continuum with energy higher than the bare bandgap. As the carriers have to thermalize before their capture by the quantum dot, increasing the excitation power is equivalent to increasing the temperature, as it leads to the increase in phonon mode population.

While we do not discount the possibility of a similar mechanism being present here, we do find that both the wurtzite lattice type of our system as well as its one-dimensional geometry suggest an alternative origin of the nonlinear linewidth broadening. Indeed, the coupling between the carriers and the phonons is expressed as $M^{ij}_{a}(\vec{k}) = A_{a}(\vec{k}) \int d^3 r \psi_{ia}^*(\vec{r}) \psi_{j,a}(\vec{r}) e^{i\vec{k}\cdot\vec{r}}$, where the parameter $A_{a}(\vec{k})$ depends on the type of coupling (deformation or piezoelectric), material parameters, and phonon energy, $\vec{k}$ is the 
phonon wavevector, and $\psi_{i,a}(\vec{r})$ is the wave function of the carrier $a$ (electron or hole) corresponding to the single-particle state $i$ (Ref.~\onlinecite{Takagahara_PRB1999,Muljarov_PRL2004,Reigue_PRL2017}). Owing to the nanowire geometry, the only phonon modes with sufficiently low energy are those propagating along the wire,
i.e., with $\vec{k}=[0,0,k]$, while the lattice vibrations in the transverse direction are quantized with energies equivalent to hundreds of Kelvin. As a result, the longitudinal phonon modes are unlikely to 
cause electronic transitions (even virtual ones) since they would need to straddle the intersubband energy gaps of hundreds of meV brought about by the small height of the quantum dot. This allows us to approximate, in the low phonon energy limit, $M^{ij}_{a}(\vec{k}) = A_{a}(\vec{k}) \delta_{i,j}$.

To date, the theoretical modeling of the linewidth broadening included the deformation potential as the only relevant mechanism of exciton-phonon coupling in zincblende materials \cite{Takagahara_PRB1999,Muljarov_PRL2004,Reigue_PRL2017}. In that mechanism, the coupling parameter for the one-dimensional phonon modes is $A_{def}(k) = \sqrt{\frac{\hbar\omega_{k}}{2\rho_M u_s^2 V}}D$,
where the phonon energy $\hbar\omega_{k}=\hbar u_s k$, $u_s$ is the speed of sound, $\rho_M$ is the mass density, $V$ is the phonon normalization volume, and $D=D_e-D_h$ is the deformation potential constant, expressed as the difference between the electron-phonon and hole-phonon coupling owing to the opposite signs of these charges. The piezoelectric coupling mechanism was found to be negligibly small, as it requires the shear strain tensor components, which are vanishingly small for longitudinal phonon modes. This is in contrast to the wurtzite crystal lattice, in which the piezoelectric effects enter through diagonal strain tensor element via the piezoelectric constant $c_{33}$. The relevant coupling parameter, adapted for the zincblende material from Ref.~\onlinecite{Takagahara_PRB1999}, has the form
$A_{piezo}(k) = -16 \pi e e_{33} \sqrt{ \frac{\hbar^2}{2\rho_M \omega_{k} V }}$, where $e$ is the electron charge. We note that the contributions from the electron-phonon and hole-phonon interactions add (rather than subtract as in the deformation coupling). Moreover, crucially, the piezoelectric coupling has (i) the opposite sign to that of the deformation mechanism, and (ii) a different dependence on the
phonon energy, $A_{piezo}(k)\propto (\omega_k)^{-1/2}$ compared to $A_{def}(k)\propto (\omega_k)^{1/2}$. As a result, the total coupling $A(k) = A_{def}(k)+A_{piezo}(k)$ is expected to depend strongly and nonlinearly on the phonon energy.

The Hamiltonian of the exciton-phonon system accounting for the approximations discussed above is:
\begin{equation}
\hat{H}_{X-ph}= E_X |X\rangle \langle X |
+ \sum_{\vec{k}} \hbar\omega_{\vec{k}} a^+_{\vec{k}} a_{\vec{k}}
+ \sum_{\vec{k}} A(|\vec{k}|) (a_{\vec{k}} + a_{\vec{k}}^+)
\end{equation}
arising because the coupling $A(|\vec{k}|)$ depends only on the magnitude of the phonon wave vector $\vec{k}=[0,0,k]$. Here, the operator $a_{\vec{k}}$ ($a_{\vec{k}}^+$) annihilates (creates) a phonon in mode $\vec{k}$. This independent boson model is exactly solvable in terms of displaced phonon operators $b_{\vec{k}} = a_{\vec{k}} + \Lambda_{\vec{k}}$ and $b_{\vec{k}}^+ = a_{\vec{k}}^+ + \Lambda_{\vec{k}}$,
where the displacement $\Lambda_{\vec{k}} = A(k) / \hbar\omega_{\vec{k}}$. Upon this transformation, our Hamiltonian is diagonal both in the excitonic and phononic degrees of freedom and reads
\begin{equation}
\hat{H}_{X-ph} = E_X |X\rangle \langle X | + 
\sum_{\vec{k}} \hbar\omega_{\vec{k}} b^+_{\vec{k}} b_{\vec{k}}
- \sum_{\vec{k}} \hbar\omega_{\vec{k}} (\Lambda_{\vec{k}})^2.
\end{equation}
The eigenvectors of our Hamiltonian can be written as the tensor products $|X\rangle \prod_{\vec{k}} |N_{\vec{k}}\rangle$, where the phonon Fock states $|N_{\vec{k}}\rangle = \frac{1}{\sqrt{N_{\vec{k}}!}} ( b^+_{\vec{k}} ) ^N |0\rangle$, and the state $|0\rangle$ is the zero-phonon state for the mode $\vec{k}$.

Upon exciton recombination, we deal with the system of non-displaced phonons described simply by the Hamiltonian $\hat{H}_{ph}=\sum_{\vec{k}} \hbar\omega_{\vec{k}} a^+_{\vec{k}} a_{\vec{k}}$ with eigenstates in the form of non-displaced phonon Fock states $|n_{\vec{k}}\rangle = \frac{1}{\sqrt{n_{\vec{k}}!}} ( a^+_{\vec{k}} ) ^n |0\rangle$. Working with the Fock phonon configurations, both in the initial and final states of the system, and considering only one phonon mode with energy $\hbar\omega_{\vec{k}}$, we can now derive the emission spectrum in the form of a series of maxima found at energies
\begin{equation}
    E(N_{\vec{k}}, n_{\vec{k}},\Lambda_{\vec{k}}) 
    = E_X - 
    \hbar\omega_{\vec{k}} (\Lambda_{\vec{k}})^2
    + \hbar\omega_{\vec{k}} (N_{\vec{k}}-n_{\vec{k}} ).
\end{equation}
The radiative transitions can be accompanied by phonon emission ($N_{\vec{k}}<n_{\vec{k}}$) or phonon absorption ($N_{\vec{k}}>n_{\vec{k}}$), placing the relevant maxima on the lower or higher side of the zero phonon line, respectively. However, owing to the very small energy scale of the relevant phonon modes, the emission spectrum will be seen as a single, broadened peak, whose intensity will be modulated by an envelope expressed in terms of the well-known Franck-Condon formula
\begin{equation}
    W(N_{\vec{k}},n_{\vec{k}},\Lambda_{\vec{k}}) =
    e^{-\Lambda_{\vec{k}}^2}
    \Lambda_{\vec{k}}^{2(n_{\vec{k}}-N_{\vec{k}})}
    \frac{N_{\vec{k}}!}{n_{\vec{k}}!} 
    \left[ L_{N_{\vec{k}}}^{n_{\vec{k}}-N_{\vec{k}}} (\Lambda_{\vec{k}}^2) \right]^2,
\end{equation}
where $n_{\vec{k}} \geq N_{\vec{k}}$ and $L_{n}^m(x)$ is the associated Laguerre polynomial (for $n_{\vec{k}} < N_{\vec{k}}$ the indices are interchanged).

The emission spectrum as a function of temperature $T$ and accounting for statistical occupation of the phonon modes is computed as a weighted superposition of the single-mode spectra. 
Upon normalization by the zero-phonon line intensity, we have:
\begin{eqnarray}
    I(\hbar\omega,T) &\propto& \sum_{\vec{k}} \frac{1}{I_0(\vec{k})}
    \sum_{N_{\vec{k}}} \sum_{n_{\vec{k}}}
    p(N_{\vec{k}},T)
    W(N_{\vec{k}},n_{\vec{k}},\Lambda_{\vec{k}})
    \nonumber\\
    &\times& \delta(\hbar\omega -E(N_{\vec{k}},
    n_{\vec{k}},\Lambda_{\vec{k}}) ),
\end{eqnarray}
with $\hbar\omega$ being the measured photon energy. Here, we consider all possible initial and final occupations of each phonon mode $\vec{k}$, weighted by the probabilities $p(N_{\vec{k}},T)=\exp\left( -\frac{N_{\vec{k}} \hbar\omega_{\vec{k}}}{k_B T}   \right) /  Z(\vec{k})$, where the statistical sum $Z(\vec{k}) =\sum_{n=0}^{\infty}\exp\left( -\frac{n \hbar\omega_{\vec{k}}}{k_B T}\right)$, and $k_B$ is the Boltzmann constant. 
The modal zero-phonon line intensity is $I_0(\vec{k})=\sum_{N_{\vec{k}}}
p(N_{\vec{k}},T) |\langle n_{\vec{k}}=N_{\vec{k}} |N_{\vec{k}}\rangle|^2$.
Our approach takes into account the fact that, calculated individually, the low occupations of any given phonon mode $\vec{k}$ are exponentially more probable than the high occupations, and therefore give more contribution to the overall emission spectrum.

We propose that the experimentally observed nonlinearity of the emission peak broadening is captured by our model -- even without the virtual excitation scheme -- in two aspects. The first, crucial aspect is the dependence of the displacement parameter $\Lambda_{\vec{k}}$ on the phonon mode energy $\hbar\omega_{\vec{k}}$. We have $\Lambda_{\vec{k}} = \Lambda_{\vec{k}}^{(def)}- \Lambda_{\vec{k}}^{(piezo)}=
\Lambda_{0}^{(def)}(\hbar\omega_{\vec{k}})^{-1/2}- \Lambda_{0}^{(piezo)}(\hbar\omega_{\vec{k}})^{-3/2}$, where $\Lambda_{0}^{(def)}$ and $\Lambda_{0}^{(piezo)}$ are energy-independent material constants describing the two coupling mechanisms, respectively. Even if the piezoelectric coupling constant is smaller than the deformation one, at sufficiently small phonon energies these two terms will cancel each other out, giving the overall displacement parameter of zero. As the material parameters for wurtzite InAs and InP are not yet known, our experimental data suggest that this cancellation occurs most likely for phonon energies of order of $0.1$ $\mu$eV, resulting in the very small emission peak broadening for low temperatures. On the other hand, at higher phonon energies the piezoelectric contribution to the parameter $\Lambda_{\vec{k}}$ decays faster than the deformation one, resulting in an effective, nonlinear increase of that parameter. As a result, the higher-energy modes, which become more populated at higher temperatures, cause the peak broadening visible at higher temperatures.

The crossover between the two regimes is systematically accounted for by the second aspect of our model, i.e., the mode occupation probabilities $p(N_{\vec{k}},T)$. As already mentioned, at low temperatures these probabilities strongly favour low-energy modes occupied with few phonons, for which the effective exciton-phonon coupling is negligibly small due to the cancellation discussed above. However, as the temperature increases, the probabilities are redistributed towards increasing occupations of more energetic phonon modes, which are more strongly coupled to the exciton. As a result, these higher modes contribute more to the emission spectrum, resulting in the nonlinear increase of the peak broadening.

We illustrate the above theoretical model with a calculation accounting for four phonon modes, with energies $\hbar\omega_{\vec{k}}=0.25$ $\mu$eV, $0.5$ $\mu$eV, $1$ $\mu$eV, and $2$ $\mu$eV. We chose the following dependence of the displacement parameter on the mode energy: $\Lambda(\varepsilon) = A_0 \left( \varepsilon^{-1/2} - \varepsilon^{-3/2}\right)$, where $\varepsilon = \hbar\omega_{\vec{k}}/\varepsilon_0$, $\varepsilon_0=0.1$ $\mu$eV, and $A_0$ was taken as a model value of $0.2$. This functional form follows directly from the subtraction of the deformation and piezoelectric contributions, and was adjusted to vanish for $\hbar\omega_{\vec{k}}=0.1$ $\mu$eV. We generated the emission spectra at a grid of photon energies with step of $2$ $\mu$eV on both sides of the zero-phonon line, but we excluded the zero phonon line itself, as its emission peak contains contributions from all phonon modes present in the system and therefore would be poorly reproduced by our model. The FWHM obtained by fitting the resulting emission peak by a Lorentzian is presented in Figure~\ref{fig_3}. Unsurprisingly, the theoretical approach,  accounting only for four modes, underestimates the line  broadening, but reproduces the experimental temperature trend qualitatively.

In summary, we have shown that it is possible to generate near-transform-limited photons from quantum dots, even when  excited above-band, by using structures that eliminate many potential broadening mechanisms. The drastic reduction in excess broadening revealed a non-linear temperature dependence that can be fully described within an independent Boson model that considers both deformation and piezoelectric coupling mechanisms, the latter expected to be more important in these wurtzite structures compared to the more typical zincblende quantum dots. 

This work was supported by Natural Sciences and Engineering Research Council of Canada


\bibliography{whiskers}   

\begin{thebibliography}{54}
\bibitem [{\citenamefont {note that since pulsed excitation{,} required for
  on-demand operation{,} is spectrally broader compared to CW excitation{,} it
  is less effective in mitigating~linewidth broadening}()}]{pulsed_vs_cw}%
  \BibitemOpen
  \bibfield  {author} {\bibinfo {author} {\bibfnamefont {We}~\bibnamefont {note
  that since pulsed excitation{,} required for on-demand operation{,} is
  spectrally broader compared to CW excitation{,} it is less effective in
  mitigating~linewidth broadening}},\ }\href@noop {} {\ }\BibitemShut {NoStop}%
\end{thebibliography}

\end{document}